\documentclass[runningheads]{llncs}
\usepackage{graphicx}
\usepackage{booktabs}
\usepackage{scrextend}

\begin{document}
\title{Deep Edge Intelligence: Architecture, Key Features, Enabling Technologies and Challenges}
\titlerunning{Deep Edge Intelligence}
 \author{Prabath Abeysekara\inst{1}\and
 Hai Dong \inst{1}\and
 A. K. Qin \inst{2}}
\institute{School of Computing Technologies, RMIT University, Melbourne, Australia
 \email{\{prabath.abeysekara, hai.dong\}@rmit.edu.au}\\
% \url{http://www.springer.com/gp/computer-science/lncs} 
\and
Computer Science and Software Engineering, Swinburne University of Technology, Hawthorn, Australia\\
\email{kqin@swin.edu.au}}
% %
\maketitle              % typeset the header of the contribution
\begin{abstract}
% With the breakthroughs in Deep Learning, recent years have witnessed the booming of Artificial Intelligence applications and services. Driven by the rapid advances in mobile computing and Internet of Things, billions of mobile and smart sensing devices are connected to the Internet, generating zettabytes of data at the network edge. The ability to imbue interconnected devices with intelligence is at the forefront of this technological revolution. In this article, we present a novel computing vision named Deep Edge Intelligence. Deep Edge Intelligence employs Deep Learning, Artificial Intelligence, Cloud and Edge Computing, 5G/6G networks, Internet of Things, Microservices, etc., aiming to provision reliable and secure intelligence services to every person and organisation at any place with low latency. The architecture, key features and enabling technologies of Deep Edge Intelligence are detailed. Finally, we reveal the primary research challenges in this area.
With the breakthroughs in Deep Learning, recent years have witnessed a massive surge in Artificial Intelligence applications and services. Meanwhile, the rapid advances in Mobile Computing and Internet of Things has also given rise to billions of mobile and smart sensing devices connected to the Internet, generating zettabytes of data at the network edge. The opportunity to combine these two domains of technologies to power interconnected devices with intelligence is likely to pave the way for a new wave of technology revolutions. Embracing this technology revolution, in this article, we present a novel computing vision named Deep Edge Intelligence (DEI). DEI employs Deep Learning, Artificial Intelligence, Cloud and Edge Computing, 5G/6G networks, Internet of Things, Microservices, etc. aiming to provision reliable and secure intelligence services to every person and organisation at any place with better user experience. The vision, system architecture, key layers and features of DEI are also detailed. Finally, we reveal the key enabling technologies and research challenges associated with it.

% In this regard, conventional machine learning techniques have rapidly adapted to various applications in multiple domains. However, DL techniques, though having demonstrated unparalleled performance primarily in Computer Vision and Natural Language Processing fields, are often subjected to significant computation and memory costs as well as massive data requirements. This poses a great challenge to empower devices at the network edge with DL capability. Nowadays, accelerated by the remarkable success of DL and IoT technologies, there is an urgent need to push the DL frontier to the network edge to fully unleash the potential values of big data. The emerging Edge Computing (EC) paradigm provides a promising way to enable this, which leverages on distributed computing concepts to push computational loads from the network core to the network edge with the aim to provide faster responses to end users. In this article, we present a novel computing framework named Deep Edge Intelligence (DEI), which is a combination of DL, AI, EC and AIoT. It enables the development and deployment of DL and AI techniques, based on EC, on edge devices, e.g., AIoT devices, where the data are generated, aiming to provide AI for every person and every organisation at any place. The key enabling technologies of DEI will be detailed. Several real-world application scenarios of DEI will be elaborated to illustrate its practical significance. Finally, we will reveal the primary research directions in this area.

\keywords{Deep Edge Intelligence \and Artificial Intelligence \and Deep Learning \and Edge Computing \and Internet of Things.}
\end{abstract}

\section{Introduction}\label{sec:intro}

The pervasive adaptation of Deep Learning (DL) driven by the recent advancements in Mobile Computing (MC) and Artificial Intelligence (AI) has opened up a myriad of opportunities across several application domains. Examples for such applications include autonomous cars, video analytics, and cognitive assistance technologies. \cite{plastiras2018edge}\cite{lin2020edge}. These applications powered by billions of mobile and Internet of Things (IoT) devices connected to the internet generate zillion bytes of data at the network edge. The opportunity to combine these two domains of technologies to power interconnected devices with intelligence is likely to pave the way for a new wave of technology revolutions. More than the technologies themselves, what lies at the forefront of this technological revolution is the vision to make the lives of people better and efficient. To realise the aforementioned vision, this data generated in exorbitant volumes at rapid velocities at the edge of the network needs to be processed and analysed to derive solutions to everyday problems of people based on DL and other AI techniques.

DL techniques in particular, having demonstrated unparalleled performance primarily in Computer Vision and Natural Language Processing fields, are often subjected to significant computation and memory costs as well as massive data requirements. This poses a great challenge to empower devices at the network edge with advanced analytics based on DL capabilities \cite{eshratifar2019towards}. In such a setting, the emerging Edge Computing (EC) paradigm provides a promising way to enable this. Sitting in close proximity to end-users and services, EC aims to provide computing, storage and network resources for applications. In other words, leveraging the distributed computing concepts to push computational loads from the network core to the network edge with the aim to provide faster responses to end users, EC provides an enticing platform to enable DL-based intelligent applications at the edge of the network. 

Recent attempts to combine EC and AI thereby fully unleashing the potential values of big data generated at the edge has led to the Edge Intelligence (EI) paradigm. EI brings together EC and AI together to shift intelligence to the edge, relieving the network infrastructure with exponentially increasing network stress. In the process, it also promises end-users with context-aware, faster intelligent services at the edge of the network. Although there has been a rapidly growing amount of research focused on enabling DL-based EI in the recent past, there is an urgent need for a more holistic framework that drives such approaches to enable true edge-native DL strategies that push the DL frontier to the network edge more comprehensively. We envision that such a holistic framework will bring together multiple key technologies to come up with a comprehensive strategy that facilitates and makes recommendations for autonomic, deeply-integrated and environment-aware, privacy-preserving, collaborative and trustworthy DL applications in an EC environment. We also hope that such an approach will bring researchers and enterprises together to outline a new class of AI strategies, algorithms and a collection of reference architectures as well as applications addressing the challenges of DL applications in the aforesaid setting.

To realise the aforementioned vision, we present a novel computing framework named Deep Edge Intelligence (DEI), which is a combination of DL, AI, EC and AIoT. DEI enables the development and deployment of DL and AI techniques, based on EC, on edge devices, e.g., AIoT devices, where the data is generated, aiming to provide AI for every person and every organisation at any place. 

% The remainder of this paper is structured as follows. Section \ref{sec:related-work} provides a comprehensive survey of existing EI approaches across multiple dimensions and outlines their limitations that led to DEI. Section \ref{sec:systems-architecture} introduces the vision of DEI and also the systems architecture of DEI. Meanwhile, Section \ref{sec:enabling-tech} comprehensively discusses the key enabling technologies of DEI. Section \ref{sec:challenges-ops} presents possible research challenges as well as opportunities. Finally, Section \ref{sec:conclusion} provides the concluding remarks.

The remainder of this paper is structured as follows. Section \ref{sec:related-work} provides a comprehensive survey of existing EI approaches. Section \ref{sec:systems-architecture} introduces the vision of DEI and also the systems architecture of DEI. Meanwhile, Section \ref{sec:enabling-tech} comprehensively discusses the key enabling technologies of DEI. Section \ref{sec:challenges-ops} presents possible research challenges as well as opportunities. Finally, Section \ref{sec:conclusion} provides the concluding remarks.

\section{Related Works}\label{sec:related-work}
% This section provides a comprehensive survey of existing edge intelligence approaches with a particular focus on the context of running deep learning-based applications. The content of this section is organised as follows

% EI for DL applications has gained increasing attention of the researchers during the recent past \cite{joshi2022enabling}. As a result, a significant number of strategies have been proposed that focus on adapting EI into many application domains. For instance, \cite{zeb2021edge} and \cite{xu2019data} proposed a deep learning based edge AI strategy for network traffic prediction and intrusion detection, respectively. Meanwhile, \cite{he2017integrated} successfully used a deep reinforcement learning based approach for integrated networking and caching in the context of vehicular networks. In other application domains, \cite{li2018deep} introduced an approach titled DeepIns for efficient manufacturing inspections whereas \cite{tian2019fog} used fog-based edge intelligence strategy in a smart home setting. In addition, there have also been attempts to use different types of deep learning approaches in an edge-native manner, as well \cite{lalapura2021recurrent}. Despite the increasing popularity and adaptation, almost all these strategies paid a lack of attention to the end-to-end aspect of running an AI strategy within a highly distributed edge environment. For instance, in such a distributed environment, manual or even semi-automatic data labelling, processing, model selection is often infeasible.

A significant number of strategies have been proposed in the recent past that focused on adapting EI into many application domains. These include video analytics, industrial IoT, cognitive assistance, smart homes, precision agriculture and trust prediction \cite{lin2020edge}\cite{abeysekara2019machine}. Out of the aforementioned strategies, some focused predominantly on using DL approaches  (e.g. Convolutional and Recurrent Neural Networks, Deep Reinforcement Learning), and other traditional optimisation techniques (e.g. Support Vector Machines), in federated and decentralised \textit{edge-native} settings \cite{lalapura2021recurrent}\cite{he2017integrated}\cite{abeysekara2020distributed}. Despite the increasing popularity and adaptation, almost all these strategies paid a lack of attention to the end-to-end aspect of running an AI strategy within a highly distributed edge environment. For instance, in such a distributed environment, manual or even semi-automatic data labelling, processing, model selection are often infeasible. 

A comprehensive proposal for running edge intelligence strategies within a 6G environment was proposed in \cite{xiao2020toward}. This work evaluates the requirements of edge AI applications, and proposes a self-learning architecture that aims to reduce the degree of human intervention across multiple aspects. These aspects include data labelling and processing, model search and consultation as well as model retraining or tuning in the face of non-stationary data. This aligns cohesively with part of our vision that aims at a more autonomous approach for edge AI in edge environments. However, it leaves out some of the most influential aspects of efficiently running an end-to-end AI strategy within an edge computing environment. These aspects include the need for self-knowledge distillation, self-organisation of knowledge sharing topologies to share knowledge amongst the participants of edge learning, self-healing in the face of failures as well as automatic hyperparameter tuning.

Meanwhile, \cite{wang2021towards} proposed an EI strategy that aims to jointly optimise multiple parameters to reduce the overall energy consumption of an edge IoT system. However, the aforementioned strategy assumes that the data processing as well as model training takes place predominantly at edge servers that are deployed in close proximity to edge devices. Therefore, it fails to cater to scenarios where learning happens on-device, and also does not take into account the need of ensuring energy efficiency of participants (e.g. mobile devices) in a more personalised or targeted manner. \cite{lim2021decentralized} proposed a dynamic resource allocation strategy for a decentralised federated learning setting while \cite{li2018edge} introduced an adaptive strategy to adaptively partition the training of a Deep Neural Network (DNN) between the edge and device to maximise resource utilisation. All the aforementioned approaches aimed to (albeit in three distinct directions) achieve deep integration between edge and device layers, if not, the edge-device synergy. However, individually, they fail to address the broader requirements of resource-efficiently running EI strategies.

\section{Vision and Systems Architecture}\label{sec:systems-architecture}

% % This section provides an elaborate description of the systems architecture of DEI. We first discuss the high-level vision and the principles that drive the proposed architecture. We, then, comprehensively detail the key layers and components that collectively embody the aforementioned systems architecture as well as their relationships to each other. 
% This section describes the vision and systems architecture of DEI. In addition, we also comprehensively detail the key layers and components that collectively embody the aforementioned systems architecture as well as their relationships to each other. 
% % To justify and compel the reader on the validity of the key design principles and their motivations, we also provide real-world examples where applicable.

% \subsection{Vision}
% \textbf{Deep Edge Intelligence} is an effort towards establishing a holistic framework that facilitates and makes recommendations for autonomic, deeply-integrated and context-aware, privacy-preserving, collaborative and trustworthy deep learning applications in an edge computing environment. It aims to bring the researchers and enterprises together to outline a new class of AI strategies, algorithms and a collection of reference architectures as well as applications addressing the challenges of deep learning applications in such a setting.

% \vspace{-1.5mm}
Below, we formally introduce the vision of DEI.
% \vspace{3mm}

\noindent\textit{\textbf{Deep Edge Intelligence} is an effort towards establishing a holistic framework that facilitates and makes recommendations for autonomic, deeply-integrated and context-aware, privacy-preserving, collaborative and trustworthy deep learning applications in an edge computing environment. It aims to bring the researchers and enterprises together to outline a new class of AI strategies, algorithms and a collection of reference architectures as well as applications addressing the challenges of deep learning applications in such a setting.}\\
% \begin{addmargin}[1em]{1em}

% \end{addmargin}
% \vspace{3mm}
% We comprehensively detail the key layers and features of the proposed systems architecture aimed at realising the aforementioned vision as well as their relationships, below. 

We detail the key elements of the proposed systems architecture aimed at realising the aforementioned vision as well as their relationships, below. 

\begin{figure*}[t]
\centering
  \includegraphics[width=\linewidth,keepaspectratio]{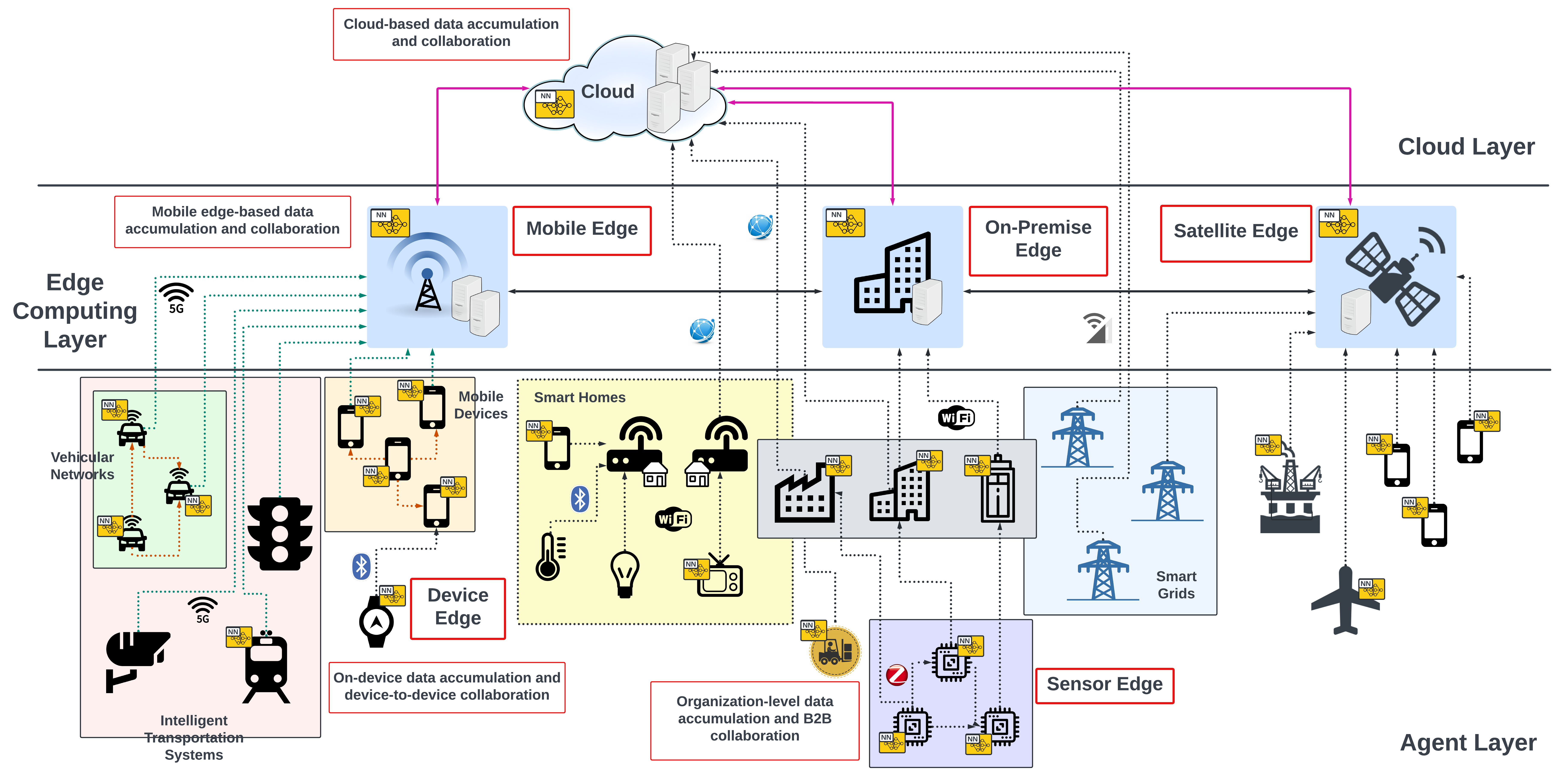}
  \caption{Systems architecture of the proposed Deep Edge Intelligence strategy from a computational perspective.}\label{fig2}
\end{figure*}

% \vspace{-3mm}
\subsection{Key Layers}
% This section outlines the key layers associated with the proposed DEI systems architecture from a computation and learning perspective. We identify three common layers associated that commonly exist in edge AI strategies, and elaborate their roles in the proposed systems architecture in the subsequent sections.
% We identify three key layers of the proposed DEI systems architecture from a computation and learning perspective, and elaborate their roles in the subsequent sections.
We identify three key layers of the proposed DEI systems architecture based on the role they perform and elaborate their characteristics based on the service scope and range, resource capacity, latency and governance and in the following sub-sections. A summary of it can also be found in Table \ref{tab:comparison-of-features}.

% \vspace{-3mm}
\subsubsection{Agent Layer}
This layer constitutes \textit{agents} that typically generate the data used for edge learning. The agents could be broadly classified based on multiple factors such as data generation strategy, data ownership, etc. Examples include mobile devices generating data of which the ownership is held by an individual, or an organisation that generates data via sensors or equipments monitoring a particular organisational process (e.g. supply-chain processes of a manufacturing plant). Furthermore, this layer also is responsible for model enforcement or use within a given application context. In addition, agent layer typically has low/medium resource capacity based on the type (e.g. mobile device, organisation, etc.), and also provides low-to-medium complexity AI services. Governance of these services, meanwhile, is either device- or organisation-specific. In scenarios where the data is generated via resource-constrained agents, the role of this layer is likely to overlap with that of the Edge Computing Layer. Since the AI services are provided within the agents themselves, this layer provides the lowest latencies for applications.

% \vspace{-3mm}
\subsubsection{Edge Computing Layer}
This layer is where the data is typically accumulated and computing takes place. Based on the type of computing infrastructure involved, we recognise various types of edge computing environments that facilitate DEI. These include device and sensor edge, mobile edge, on-premise or enterprise edge, drone edge and satellite edge, etc. Typically, the Edge Computing (EC) layer maintains either one-way or two-way communication between the agent and cloud layers via various communication infrastructure such as 4G LTE, 5G, WAN/LAN, Bluetooth, backhaul links as well as other open and proprietary communication protocols. Edge Computing layer typically carries medium resource capacity compared to the other layers, and also provide AI services of medium complexity and range to users and applications. Also, service governance is generally enforced via either zoned or partitioned manner (e.g. per individual or group of EC environments within an EC topology). Since devices access the AI services provided by this layer via some low-range communication medium described above, the latency incurred is greater than that of the agent layer and yet, significantly lower than the services in the Cloud layer.

% \vspace{-3mm}
\subsubsection{Cloud Layer}
Sitting at the top of the proposed system architecture, the cloud layer plays a vital role in realising the goals of DEI. For instance, the virtually infinite computing and storage resources available within the cloud infrastructure allow running more advanced and data-intensive AI strategies at scale. The cloud also plays the role of a knowledge repository and aggregator for all or some edge AI models trained in distributed manner to facilitate Collaborative Intelligence. In addition, it helps derive globally optimal resource allocation strategies for edge AI applications and services in order to ensure they adhere to the announced QoS specifications. The cloud also acts as a repository for services that complement those provided at the edge. Such a centralised service repository allows service consumers to look up alternative services in scenarios where suitable services are not present in the edge layer. The other benefits include the ability to enforce centralised governance over how AI services are provisioned, distributed and used. Due to the number of hops a user request has to travel through, this layer exhibits the highest latencies on AI service access.

\begin{table}[h!]
\centering
% \resizebox{\textwidth}{!}{
\begin{tabular*}{\textwidth}{@{\extracolsep{\fill}\quad}lccc}
    \toprule
    \textbf{Focus Area} & \textbf{Agent Layer} & \begin{tabular}{@{}c@{}}\textbf{Edge Computing} \\ \textbf{Layer}\end{tabular} & \textbf{Cloud Layer}\\\midrule
    Service scope and range & Low & Medium & High\\\midrule
    Resource capacity & Low/Medium & Medium & High\\\midrule
    Latency & Low & Medium & High\\\midrule
    Service Governance & \begin{tabular}{@{}c@{}} On-device or \\ Organisation-based\end{tabular} & \begin{tabular}{@{}c@{}} Zoned or \\ Partitioned\end{tabular} & Fully centralised\\\midrule
\end{tabular*}
% }
% \vspace{1mm}
\caption{Comparison of the key layers against the degree and scope of their capabilities.}
\label{tab:comparison-of-features}
% \vspace{-3mm}
\end{table}

\subsection{Key Features}
% Below, we explain the key features of the proposed system architecture of DEI for completeness.
Below, we explain the key features of the proposed system architecture of DEI for completeness.

% \vspace{-3mm}
\subsubsection{Autonomous Execution} 
Designing and developing an end-to-end AI strategy for applications running at the edge is a laborious task due to multiple reasons. For instance, in a typical edge-based application, the data is accumulated in a distributed manner across diverse environments varying from personal devices to datacentres managed by enterprises. Therefore, many phases of AI strategies that have traditionally been performed either manually or semi-automatically under the supervision of domain experts have now become infeasible in such a setting. Examples include (not limited to), data labelling and annotation, hyperparameter tuning, knowledge aggregation, failure recovery. DEI aims to take an autonomic approach to carry out the aforementioned tasks in edge-based intelligent applications. To that end, DEI puts a stronger emphasis on self-learning (e.g. self-supervised learning), self-tuning and self-organising networks for knowledge sharing as well as self-healing approaches to derive more autonomic strategies for edge-based applications.

% \vspace{-3mm}
\subsubsection{Deep Integration and Context-Awareness}
Edge computing environments typically operate under heterogeneous operating conditions (e.g. network bandwidth, computing and storage resources, energy utilisation) due to their inherently distributed systems architecture and other factors such as uneven distribution of computing, storage and network resources. This gives rise to multiple \textit{context-environments} within edge computing topologies \cite{abeysekara2021data}. In addition, the distributions of data, accumulated in different scales based on the type of edge environment (e.g. mobile devices, organisations) may also exhibit varying characteristics owing to the existence of such context environments. Therefore, the deep integration of such environmental and operational conditions extracted using network statistics, QoS monitoring systems as well as varying data characteristics into DEI applications is paramount in order to provide accurate and predictable application utility to the users. To that end, DEI motivates intelligent applications running at the edge to be context-aware and employ intelligent strategies for edge resource provisioning, ensuring energy efficiency, etc.

% \vspace{-3mm}
\subsubsection{Collaborative Intelligence}
In an EI setting, the actors involved in training distributed AI models only see a split view of the world irrespective of the type of learning architecture used (e.g. cross-device or cross-siloed distributed learning). Such a phenomenon driven predominantly by the inherent distributed architecture of edge computing, often brings significant challenges in setting up good quality predictive models. On the one hand, the actors operating at the edge (e.g. mobile phones or organisations) accumulate data corresponding to specific activities they are engaged in. This can potentially introduce personal or organisational biases, which can hamper the quality of the learnt models due to overfitting. Furthermore, there could also be unevenness of the data distribution, or even noises or anomalous data annotations across different actors contributing to the same issues outlined above. Collaborative Intelligence, in such a context, provides an opportunity for actors that are either homogeneous (i.e. from the same domain of knowledge) or heterogeneous (i.e. from different domains of knowledge) to either borrow or share knowledge from each other to derive better generalisable models. To that end, DEI, at the heart of its design, also encourages knowledge sharing amongst applications via collaboration.

% \vspace{-3mm}
\subsubsection{Privacy Preservation}
An AI strategy for edge applications brings together many systems and actors in various layers and capacities. The natural expectation is that these systems may behave within their expected boundaries of execution and actors will act rationally and learn only the bare minimum amount of information about the users enabling them to perform their role within the underlying quality assurance framework. Although the aforementioned aspects sound obvious and straightforward, putting together a systems framework to ensure their adherence is oftentimes a daunting task. This is particularly true for edge AI applications where the actors (e.g. mobile devices) collaborating together and systems they interact with are heterogeneous and disparate. Therefore, DEI regards user privacy preservation as a key decision principle in its systems architecture rather than an afterthought, as it is often observed in many real-world scenarios. Furthermore, DEI also aims to drive more pragmatic solutions adhering the various privacy laws governing how the user information is collected, processed and used such as General Data Protection Regulation (GDPR), Health Insurance Portability and Accountability Act (HIPAA) and Australian Privacy Principles (APPs), etc., as well.

% \vspace{-3mm}
\subsubsection{Trustworthiness}
EI applications are often prone to various adversarial and non-adversarial related challenges at different phases of their life cycles due to their inherently distributed systems architecture and other factors. These affected phases include data accumulation, pre-processing, model training, evaluation as well as serving inferences. Such a phenomenon often hinders the ability of these applications to provide trustworthy outcomes that are inline with the expectations of the users. From an adversarial perspective, the potential limitations and vulnerabilities include (not limited to), the lack of central authority, heterogeneity of the systems and actors involved, often weak encryption and encoding schemes used in devices in order to maintain affordability and compensate for low configurations. On the other hand, the non-adversarial aspects challenging the edge AI applications include the lack of quality in data, unreliable agents or actors, poorly provisioned infrastructure for serving inferences, biases that exist in raw data and trained models, etc. Therefore, DEI also focuses on a range of such aspects to establish and maintain the end-to-end trustworthiness of edge AI applications in order to improve user acceptance towards them.

% \vspace{-4mm}
\section{Enabling Technologies}\label{sec:enabling-tech}
% \vspace{-2mm}
This section discusses several technologies (including, but not limited to) that could potentially play a pivotal role in enabling DEI realise its goals. For clarity, we have broadly classified them into multiple categories in the form of, AI Technologies, Service Architectures and Infrastructures as well as Communication Protocols and Network Infrastructures.

% \vspace{-3mm}
\subsection{AI Technologies}
\subsubsection{Federated Learning (FL) and Distributed Machine Learning (DML)}
FL and DML are two fundamental pillars on which DEI relies on. OUt of them, FL predominantly aims to collaboratively train a shared predictive model amongst a group distributed \textit{agents} via central coordination \cite{kairouz2021advances}. In the process, it allows harnessing the computing resources available within these agents to achieve such an objective without requiring any movement of the raw data outside the underlying agents. Meanwhile, DML in a DEI setting may focus on parallelising the training of a predictive model amongst multiple disparate and typically homogeneous agents \cite{chen2021distributed}. Recent variants in FL and DML that emerged such as Federated Transfer Learning, Personalised Federated Learning as well as Split Learning presented a wealth of opportunities to achieve collaborative intelligence as well as privacy preservation, which lie at the heart of DEI.

% \vspace{-3mm}
\subsubsection{Self-Supervised Learning (SSL)} 
SSL provides an potent framework to learn unsupervised representations in the face of unlabelled data. It also allows data-efficiently generating predictive models with higher generalisability in the domain of typically data-hungry deep learning algorithms even in the midst of incomplete, transformed or corrupt data \cite{liu2021self}. The aforementioned characteristics show immense potential and promise versatility of SSL across various DEI applications ranging from those that collect comparatively small (e.g. language modelling on mobile devices) to large volumes of data (e.g. camera sensor data generated by autonomous vehicles). In addition, the ability of SSL approaches to harness representations from unlabelled data also allows autonomously training predictive models in highly distributed settings as that of DEI applications where expert intervention is infeasible. Therefore, it could be deemed an integral enabler that could potentially help realising the vision of DEI.

% \vspace{-3mm}
\subsubsection{Swarm Intelligence (SI)} 
This class of bio-inspired algorithms allows a group of decentralised and homogeneous agents to collectively derive a global solution to a given problem by interacting with their peers as well as environments. SI algorithms foster self-learning as well as adaptability, and often find utility in efficiently solving non-linear objectives towards an often globally optimal solution within reasonable time limits \cite{chakraborty2017swarm}. To that end, they could be deemed vastly relevant in the context of allowing collaborative intelligence amongst the agents participating in DEI applications. Notable variants include Particle Swarm Optimisation, Ant Colony Optimisation, and Cuckoo Search \cite{chakraborty2017swarm}.

% \vspace{-3mm}
\subsubsection{Automated Machine Learning (AutoML)}
AutoML predominantly aims to automate the entire process of rolling out AI applications, as its name implies, with a significantly less amount of involvement from the human experts. Reaching beyond the automation of merely the process of generation of an AI model, it also facilitates the automation of activities from data preparation, feature engineering, hyperparameter optimisation (HPO), model generation, and model evaluation, which had traditionally required significant amount of expert knowledge and intervention \cite{he2021automl}. Considering the complexity of DEI applications from distributed agents training predictive models to added layers of complexity such as the need to autonomously select participants for CI, run HPO across massively distributed agents as well as the continuous retrain and fine tune the models, AutoML plays an indispensable element in the vision of DEI.

% \vspace{-3mm}
\subsubsection{Privacy Preserving Machine Learning (PrivateML)}
PrivateML embodies efforts to incorporate anonymisation techniques into existing AI strategies or developing novel approaches to preserve data privacy, and by extension, user privacy, of AI applications \cite{xu2021privacy}. Existing work on FL proposing to keep the raw data close to where it originates and share only the parameters of the trained predictive models for CI has gained attraction in recent times as a mode of PrivateML in EI applications. In addition, attempts to enforce more stringent privacy preservation based on differential privacy \cite{gupta2018distributed}\cite{thapa2022splitfed} have shown great promise in terms of catering to the requirements of DEI outlined previously. In the meantime, recently proposed frameworks such as PGU, which stands for \textit{Phase-Guarantee-Technical Utility} can also be deemed relevant in the context of systematically understanding or evaluating different PrivateML approaches \cite{xu2021privacy}. Such approaches can allow looking at existing PrivateML strategies from a holistic perspective and determine suitable strategies that cater to the inherent requirements of DEI applications.

\subsection{Service Architectures and Infrastructures}

\subsubsection{Microservices}
Popularly adapted for cloud applications, microservices can be presented as a key enabling service architecture pattern driving the future DEI applications. For instance, typical DEI applications tend to harness the power of data accumulated across heterogeneous data sources ranging from mobile devices to edge and cloud platforms. This data varies in its structure, volume and velocity at which it is accumulated. In such a heterogeneous setting, microservices allows exposing intelligent, granular, data-driven services as programmable functional components adhering the systems characteristics of edge computing systems \cite{al2022ai}. In addition, the inherently distributed characteristics of edge environments challenge DEI application developers from a perspective of resource allocation and management given the heterogeneous computing and storage resources available in different edge nodes. In such an environment, microservices enable transparently allocate resources in a serverless manner and allow application developers to focus predominantly on the business requirements while also allowing serving trained predictive models scalably to interested consumers. Furthermore, mobility of services and their consumers in DEI applications demand services to be sporadically provisioned, decommissioned and migrated across edge environments \cite{chen2022dynamic}. The modularity allowed by the microservices architecture greatly simplifies all three aforementioned aspects.

% \vspace{-3mm}
\subsubsection{Container Technologies}
Containers, and in particular, \textit{edge containers} allow decentralising the application components and services to the edge of the network \cite{al2022ai}. Therefore, such an edge fabric powered by container platforms such as Docker\footnote{https://www.docker.com/} \cite{al2022ai} and Kubernetes\footnote{https://kubernetes.io/} \cite{rausch2019towards}, enable reducing network costs as well as response times thereby facilitating delay-sensitive DEI applications. Furthermore, not only do they help realise the promises of the microservices architecture for DEI applications, such container technologies also facilitate the easy orchestration or choreography of microservices as appropriate \cite{al2022ai}. Amongst other benefits, the portability allowed by containers also make the scalable deployment, provisioning and migration of DEI application components and services seamless irrespective of the underlying computing architectures, etc.

% \vspace{-3mm}
\subsection{Communication Protocols and Network Infrastructures}
% \vspace{-2mm}
\subsubsection{5G and 6G}
Efficient communication infrastructure powered by faster communication protocols such as 5G is integral to the realisation of the promised goals of DEI. To that end, 5G provides significantly low latencies for delay-sensitive DEI applications as well as higher capacity compared to older-generation networking infrastructures. Apart from that, the promised additions such as virtual network slicing in 5G could potentially allow DEI application providers to cater to provide more stringent and predictable QoS guarantees to their consumers \cite{wu2022ai}. Meanwhile, 6G, despite still being conceptualised, already promises to provide \textit{AI-native} networking capabilities to future AI applications supporting potentially microsecond latency with significantly higher network capacities. In addition, it is also believed to offer several advanced and novel features such as context-awareness, self-aggregation, re-configurable intelligent network surfaces, etc. \cite{dang2020should} which are likely to help provide consumers of DEI applications with richer user experience.

% \vspace{-3mm}
\subsubsection{Edge Infrastructures}\label{subsub:edge-infra}
DEI relies on and intends to consolidate resources available across several edge infrastructures as relevant for edge-based AI applications. In other words, a particular DEI application can potentially use either individual or multiple edge platforms in collaboration with cloud infrastructures to realise their goals. We identify the following (including, but not limited to) edge infrastructures, which could power future DEI applications.

\begin{itemize}
    \item \textit{Mobile Edge} \cite{beck2014mobile}: Co-existing within Radio Access Networks (RANs) of mobile network providers, the mobile edge provides pooled computing and storage resources to smart IoT devices and services in close proximity.
    \item \textit{Device/Sensor Edge} \cite{kairouz2021advances}: Provides the ability to train and use predictive models \textit{on-device} atop locally-accumulated data without needing to transmit them into an outside computing environment.
    \item \textit{Drone Edge} \cite{alsamhi2021drones}: Acting as relay stations to nearby resource-constrained IoT devices, drone edge utilises drones as intelligent edge nodes to aid data accumulation while providing efficient computing capabilities for AI model training.
    \item \textit{Satellite Edge} \cite{zhang2019satellite}: allows devices without any nearby mobile edge computing (MEC) environments to use edge computing services via satellite links.
    \item \textit{On-premise Edge} \cite{padmanabhan2021towards}: Refers to edge gateways, computing environments established within manufacturing plants, etc. to accumulate data and train predictive models within a particular organisational context.
\end{itemize}

% \section{Real-world DEI Applications}\label{sec:apps}
% For completeness, we briefly discuss some concrete applications that benefit from the DEI paradigm, below.

% \subsection{Intelligent Transportation Systems}
% \subsection{Smart Healthcare Systems}
% \subsection{Smart Grids, Intelligent Energy Aggregation and Distribution}

% \vspace{-4mm}
\section{Research Challenges and Opportunities}\label{sec:challenges-ops}
% \vspace{-2mm}
Herein we discuss some of the challenges in realising the vision of DEI and some of the exciting opportunities that lie ahead.

% \vspace{-3mm}
\subsubsection{Autonomous Execution and Evolution}
There is a growing need to automate several aspects of DL pipelines within a DEI setting. Examples include data labelling and annotation, model search and construction, hyperparameter tuning, knowledge aggregation, failure recovery, etc. These requirements bring in several opportunities in the context of designing algorithms and strategies that enable self-learning and self-tuning as well as self-organising and self-healing approaches for CI. In particular, two directions that have received little attention are the self-tuning of DL models and self-healing in the context of CI strategies running at the edge. From a self-tuning perspective, there is a demand for novel approaches that take into account the inherently distributed nature of DEI applications as well as other aspects such as communication efficiency. Meanwhile, from a self-healing perspective, most approaches in the current literature proposed synchronous algorithms that are likely to fail due to one or more failed agents. Therefore, the applicability of asynchronous algorithms and frameworks that are more fault tolerant need to be explored. Furthermore, in most practical scenarios the characteristics or statistical properties of data could change (i.e. \textit{concept-drift}) as well as application requirements evolve over time. As a result, DL strategies that autonomously evolve (e.g. Online Deep Learning \cite{sahoo2017online}) in response to such dynamic requirements need to be investigated, as well. This leads us to opportunities on efficient approaches that can cost-effectively fine-tune or re-train DL models to cater to such requirements.

% \vspace{-3mm}
\subsubsection{Efficient Collaborative Intelligence}
CI, which is a key pillar of DEI as introduced previously, relies heavily on the effectiveness of \textit{how} the agents collaborating to share knowledge with each other. To that end, research enabling such efficient CI can be broadly classified into two categories as approaches that enable \textit{cross-agent collaboration} and those that facilitate \textit{cross-layer collaboration}. The first category includes identifying different types of knowledge sharing and aggregation requirements of applications; designing algorithms and frameworks that are not merely problem- or application-specific, but are relevant to many common or related classes of problems; self-organising approaches to formulate knowledge sharing or aggregation topologies of different structures (e.g. hierarchical, hub-and-spoke) based on various incentive strategies (e.g. to achieve fairness in model accuracy across different local models, improve the accuracy of a single federated model, etc.). On the other hand, the second category includes approaches that allow efficient splitting of DL layers amongst the agent, edge and cloud layers, designing feature compression strategies to minimise the network stress on the networking infrastructure, etc.

% \vspace{-3mm}
\subsubsection{Dynamic Edge Resource Allocation and Optimisation}
Edge computing environments typically provide comparatively less computing, storage and network resources for applications at the network edge, compared to cloud environments. The resulting competition for resources by DEI applications, could drive the exploration of novel strategies that are more tight-knit with the underlying environment factors to provide dynamic and adaptive solutions. In other words, we anticipate novel research focused on dynamic resource allocation strategies that jointly and simultaneously optimise models while allocating resources such as network bandwidth; computing and storage requirements; energy efficiency, etc through the efficient integration of systems and networks statistics, monitoring data available at the edge \cite{hu2015mobile}.

% \vspace{-3mm}
\subsubsection{Privacy Preservation}
Sharing a similar sentiment with researchers in PrivateML, we envision the requirements for further research on open problems such as providing satisfactory privacy guarantees while maintaining the predictive power of the trained DL strategies, designing measures to ensure communication-efficiency and counter the effect of bloated data caused by encoding and encryption schemes, ensuring fairness as well as robustness while enabling privacy preservation, etc. \cite{xu2021privacy}. In addition, some PrivateML techniques using data anonymisation could potentially interrupt the ability of the underlying DL strategy to allow CI. Therefore, new research avenues that investigate combining both CI and PrivateML would also be required. Other potential directions would include designing more computationally-efficient strategies, allowing adaptive and autonomous ways to enable privacy preservation while ensuring model utility via means of introducing privacy perturbation budgets as well as enabling interoperability \cite{xu2021privacy}.

% \vspace{-3mm}
\subsubsection{Ensuring Trustworthiness}

DEI may attract many adversarial and non-adversarial challenges. Therefore, from an adversarial perspective, it is imperative to design and develop adaptive and data-driven methods that can assess, detect and act against adversarial attacks in a DEI setting \cite{xiao2020toward}. These methods include holistic approaches to threat modelling taking into account attacks covering the entire spectrum of tasks associated with a DL pipeline such as data accumulation, pre-processing, model training, evaluation as well as inference. Examples for such attacks include data poisoning and evasion, model theft, logic corruption, etc \cite{benzaid2020ai}. In addition, it is equally important to design and develop strategies to evaluate the threat landscapes not only from a DL or data perspective, but also from an end-to-end systems perspective including the underlying systems and network infrastructure, secure data communication, etc. Other potential opportunities could lie in the domain of developing decentralised authentication and authorisation protocols, trust-based security, light-weight and computationally-efficient yet reasonably rich encoding and encryption schemes, etc. Meanwhile, from a non-adversarial perspective, possible future research opportunities include investigating approaches that can counter the effect of the lack of quality in data, unreliable actors, poorly provisioned infrastructure for serving trained models, biases that exist in raw data as well as trained models, etc.

% \vspace{-3mm}
\subsubsection{Evaluation}
At present, the lack of real-world and public datasets has made it challenging to model and evaluate a given DEI strategy with a reasonable amount of confidence that it fully solves the underlying problem. Although there have been efforts to develop simulators such as \cite{blakley2020simulating}, these only cover either a smaller subset of edge environments or application scenarios. In addition, the systems and network architectures of several emerging types of edge environments outlined in Section \ref{subsub:edge-infra} themselves are still mostly experimental. Therefore, we envision a significant amount of effort will go into in the near future on conversing and modelling such systems architectures as well as generating tools and publicly-accessible datasets to allow researchers come up with more meaningful research in the domain of DEI.

% \vspace{-2.5mm}
\section{Conclusion}\label{sec:conclusion}
This article introduced a novel paradigm \textit{Deep Edge Intelligence} (DEI) that brings together the best of Deep Learning (DL), Artificial Intelligence (AI), Edge Computing (EC) and Artificial Intelligence of Things (AIoT) to come up with holistic framework for designing DL-based applications on the network edge. We also discussed the systems architecture of DEI including its key layers and features, as well as elaborated the vision behind it. Furthermore, we also provided comprehensive details on potential enabling technologies that can make DEI a reality, some example real-world applications as well as research challenges and opportunities that lie ahead. Our hope is that this work will spark further interest amongst a wide audience of researchers and pave the way for new research directions into more holistic and pragmatic adaptation of DL in edge environments.

\section*{Acknowledgement}
This research was supported by the Australian Government through the Australian Research Council’s Discovery
Projects funding scheme (project DP220101823). The views
expressed herein are those of the authors and are not necessarily those of the Australian Government or Australian
Research Council.

% \vspace{-2mm}
\bibliographystyle{splncs04}
\bibliography{dei-2022.bib}

\begin{thebibliography}{10}
\providecommand{\url}[1]{\texttt{#1}}
\providecommand{\urlprefix}{URL }
\providecommand{\doi}[1]{https://doi.org/#1}

\bibitem{plastiras2018edge}
Plastiras, G., Terzi, M., Kyrkou, C., Theocharidcs, T.: Edge intelligence:
  Challenges and opportunities of near-sensor machine learning applications.
  In: ASAP 2018. pp.~1--7. IEEE (2018)

\bibitem{lin2020edge}
Lin, S., Zhou, Z., Zhang, Z., Chen, X., Zhang, J.: Edge intelligence in the
  making: optimization, deep learning, and applications. SYN LECT LEARN NETW
  ALGORITHMS  \textbf{1}(2),  1--233 (2020)

\bibitem{eshratifar2019towards}
Eshratifar, A.E., Esmaili, A., Pedram, M.: Towards collaborative intelligence
  friendly architectures for deep learning. In: ISQED 2019. pp. 14--19. IEEE
  (2019)

\bibitem{abeysekara2019machine}
Abeysekara, P., Dong, H., Qin, A.K.: Machine learning-driven trust prediction
  for mec-based iot services. In: ICWS 2019. pp. 188--192. IEEE (2019)

\bibitem{lalapura2021recurrent}
Lalapura, V.S., Amudha, J., Satheesh, H.S.: Recurrent neural networks for edge
  intelligence: a survey. ACM COMPUT SURV  \textbf{54}(4),  1--38 (2021)

\bibitem{he2017integrated}
He, Y., Zhao, N., Yin, H.: Integrated networking, caching, and computing for
  connected vehicles: A deep reinforcement learning approach. IEEE T VEH
  TECHNOL  \textbf{67}(1),  44--55 (2017)

\bibitem{abeysekara2020distributed}
Abeysekara, P., Dong, H., Qin, A.K.: Distributed machine learning for
  predictive analytics in mobile edge computing based iot environments. In:
  IJCNN 2020. pp.~1--8. IEEE (2020)

\bibitem{xiao2020toward}
Xiao, Y., Shi, G., Li, Y., Saad, W., Poor, H.V.: Toward self-learning edge
  intelligence in 6g. IEEE COMMUN MAG  \textbf{58}(12),  34--40 (2020)

\bibitem{wang2021towards}
Wang, Q., Xiao, Y., Zhu, H., Sun, Z., Li, Y., Ge, X.: Towards energy-efficient
  federated edge intelligence for iot networks. In: ICDCSW 2021. pp. 55--62.
  IEEE (2021)

\bibitem{lim2021decentralized}
Lim, W.Y.B., et~al.: Decentralized edge intelligence: A dynamic resource
  allocation framework for hierarchical federated learning. IEEE T PARALL DISTR
   \textbf{33}(3),  536--550 (2021)

\bibitem{li2018edge}
Li, E., Zhou, Z., Chen, X.: Edge intelligence: On-demand deep learning model
  co-inference with device-edge synergy. In: MECOMM'18. pp. 31--36 (2018)

\bibitem{abeysekara2021data}
Abeysekara, P., Dong, H., Qin, A.: Data-driven trust prediction in mobile edge
  computing-based iot systems. IEEE T SERV COMPUT  (2021)

\bibitem{kairouz2021advances}
Kairouz, P., et~al.: Advances and open problems in federated learning. FOUND
  TRENDS in MACH LEARN  \textbf{14}(1--2),  1--210 (2021)

\bibitem{chen2021distributed}
Chen, M., et~al.: Distributed learning in wireless networks: Recent progress
  and future challenges. IEEE J SEL AREA COMM  (2021)

\bibitem{liu2021self}
Liu, X., Zhang, F., Hou, Z., Mian, L., Wang, Z., Zhang, J., Tang, J.:
  Self-supervised learning: Generative or contrastive. IEEE T KNOWL DATA EN
  (2021)

\bibitem{chakraborty2017swarm}
Chakraborty, A., Kar, A.K.: Swarm intelligence: A review of algorithms. LECT
  NOTES ARTIF INT pp. 475--494 (2017)

\bibitem{he2021automl}
He, X., Zhao, K., Chu, X.: Automl: A survey of the state-of-the-art.
  KNOWL-BASED SYST  \textbf{212},  106622 (2021)

\bibitem{xu2021privacy}
Xu, R., Baracaldo, N., Joshi, J.: Privacy-preserving machine learning: Methods,
  challenges and directions. arXiv preprint arXiv:2108.04417  (2021)

\bibitem{gupta2018distributed}
Gupta, O., Raskar, R.: Distributed learning of deep neural network over
  multiple agents. J NETW COMPUT APPL  \textbf{116}, ~1--8 (2018)

\bibitem{thapa2022splitfed}
Thapa, C., et~al.: Splitfed: When federated learning meets split learning. In:
  AAAI 2022. vol.~36, pp. 8485--8493 (2022)

\bibitem{al2022ai}
Al-Doghman, F., Moustafa, N., Khalil, I., Tari, Z., Zomaya, A.: Ai-enabled
  secure microservices in edge computing: Opportunities and challenges. IEEE T
  SERV COMPUT  (2022)

\bibitem{chen2022dynamic}
Chen, X., et~al.: Dynamic service migration and request routing for
  microservice in multi-cell mobile edge computing. IEEE INTERNET THINGS
  (2022)

\bibitem{rausch2019towards}
Rausch, T., et~al.: Towards a serverless platform for edge $\{$AI$\}$. In:
  HotEdge 19 (2019)

\bibitem{wu2022ai}
Wu, W., Zhou, C., Li, M., Wu, H., Zhou, H., Zhang, N., Shen, X.S., Zhuang, W.:
  Ai-native network slicing for 6g networks. IEEE WIREL COMMUN  \textbf{29}(1),
   96--103 (2022)

\bibitem{dang2020should}
Dang, S., Amin, O., Shihada, B., Alouini, M.S.: What should 6g be? NAT ELECTRON
   \textbf{3}(1),  20--29 (2020)

\bibitem{beck2014mobile}
Beck, M.T., Werner, M., Feld, S., Schimper, S.: Mobile edge computing: A
  taxonomy. In: AFIN-IARIA 2022. pp. 48--55. Citeseer (2014)

\bibitem{alsamhi2021drones}
Alsamhi, S.H., et~al.: Drones’ edge intelligence over smart environments in
  b5g: blockchain and federated learning synergy. IEEE T GREEN COMMUN and NETW
  \textbf{6}(1),  295--312 (2021)

\bibitem{zhang2019satellite}
Zhang, Z., Zhang, W., Tseng, F.H.: Satellite mobile edge computing: Improving
  qos of high-speed satellite-terrestrial networks using edge computing
  techniques. IEEE NETW  \textbf{33}(1),  70--76 (2019)

\bibitem{padmanabhan2021towards}
Padmanabhan, A., et~al.: Towards memory-efficient inference in edge video
  analytics. In: HotEdgeVideo 2021. pp. 31--37 (2021)

\bibitem{sahoo2017online}
Sahoo, D., Pham, Q., Lu, J., Hoi, S.C.: Online deep learning: Learning deep
  neural networks on the fly. arXiv preprint arXiv:1711.03705  (2017)

\bibitem{hu2015mobile}
Hu, Y.C., Patel, M., Sabella, D., Sprecher, N., Young, V.: Mobile edge
  computing—a key technology towards 5g. ETSI white paper  \textbf{11}(11),
  1--16 (2015)

\bibitem{benzaid2020ai}
Benza{\"\i}d, C., Taleb, T.: Ai for beyond 5g networks: a cyber-security
  defense or offense enabler? IEEE NETW  \textbf{34}(6),  140--147 (2020)

\bibitem{blakley2020simulating}
Blakley, J.R., Iyengar, R., Roy, M.: Simulating edge computing environments to
  optimize application experience. School of Computer Science Carnegie Mellon
  University, Technical Report CMU-CS-20-135  (2020)

\end{thebibliography}

\end{document}